\documentclass[12pt,preprint]{aastex}
\begin{document}

\title {NGC 5253 and ESO269--G058: Dwarf Galaxies With a Past 
\altaffilmark{1} \altaffilmark{2}}

\author{T. J. Davidge}

\affil{Herzberg Institute of Astrophysics,
\\National Research Council of Canada, 5071 West Saanich Road,
\\Victoria, B.C. Canada V9E 2E7\\ {\it email: tim.davidge@nrc.ca}}

\altaffiltext{1}{Based on observations obtained at the
Gemini Observatory, which is operated by the Association of Universities
for Research in Astronomy, Inc., under a co-operative agreement with the
NSF on behalf of the Gemini partnership: the National Science Foundation
(United States), the Particle Physics and Astronomy Research Council
(United Kingdom), the National Research Council of Canada (Canada),
CONICYT (Chile), the Australian Research Council (Australia), CNPq (Brazil),
and CONICET (Argentina).}

\altaffiltext{2}{This publication makes use of data products
from the Two Micron All Sky Survey, which is a joint project of the University of
Massachusetts and the Infrared Processing and Analysis Center/California
Institute of Technology, funded by the National Aeronautics and Space Administration
and the National Science Foundation.}

\begin{abstract}

	Deep $r'$ and $i'$ images obtained with GMOS on Gemini South are used to 
probe the bright stellar content in the outer regions of the 
Centaurus Group dwarf galaxies NGC 5253 and ESO269--G058. Red giant branch (RGB) stars 
are traced out to a distance of 8 kpc along the major axis of NGC 5253, and 6 kpc in 
ESO269--G058. The outer regions of both galaxies are metal-poor; 
RGB stars located between projected major axis distances of 2 and 4 kpc in NGC 5253 have 
[Fe/H] $\sim -1$, whereas RGB stars in the corresponding portion of ESO269--G058 have 
[Fe/H] $\sim -1.8$. Stars with metallicities that differ from the mean by more 
than a few tenths of a dex make only a modest contribution to the stellar content in 
the outer regions of both galaxies. A population of bright asymptotic giant 
branch (AGB) stars is seen in both galaxies. In the particular case of NGC 5253, roughly 
1 -- 10\% of the stellar mass may have formed during the past 
few hundred million years, and it is suggested that the progenitors of the 
two recent SN Ia in this galaxy may have formed at this time. That NGC 5253 has 
experienced either episodic or continuing elevated levels of star formation 
during the past few hundred million years is reminiscent of what is seen in 
other dwarf starburst galaxies, such as NGC 3077, and it is argued that the current 
episodes of star formation in NGC 5253 and ESO269--G058 may have been triggered up 
to $\sim 1$ Gyr in the past. Finally, a distance modulus is computed for each galaxy 
based on the brightness of the RGB-tip, and the results are $\mu_0 = 
27.48 \pm 0.14$ for NGC 5253, and $27.93 \pm 0.18$ for ESO269--G058.

\end{abstract}

\keywords{galaxies: individual (NGC 5253) - galaxies: individual (ESO269-G058) - 
galaxies: dwarf - galaxies: stellar content - galaxies: evolution - galaxies: 
distances and redshifts}

\section{INTRODUCTION}

	Star formation in the early Universe likely started in gas-rich protogalactic 
fragments, which may have been the birth place of globular clusters (Kravtsov \& 
Gnedin 2005). These fragments subsequently merged to form progressively larger systems. 
The incidence of merger activity peaked in the distant past, and simulations 
predict that the assembly of the Galaxy was largely complete at least 9 
Gyr ago (e.g. Bullock \& Johnston 2005). The original companions of the 
Galaxy are thought to have been completely assimilated, 
although fossil remnants may survive to the present day in the Solar neighborhood
(e.g. Helmi et al. 2006). The present-day Galactic satellites 
were likely accreted only during the past few Gyr (Bullock \& 
Johnston 2005), and have probably not yet contributed to the Galactic stellar content 
(Abadi, Navarro, \& Steinmetz 2006).

	Although the epoch of large-scale galaxy assembly occured far in the past, 
nearby dwarf galaxies with elevated star formation rates (SFRs) may provide glimpses of 
conditions in the earliest systems, as well as insight into the timescale of 
satellite accretion in the local Universe. While having a 
total mass that is similar to the Local Group (Karachentsev 
et al. 2002), the Centaurus group contains galaxies that have no obvious Local Group 
counterpart (e.g. NGC 5128; M83), making Centaurus an interesting target for 
studies of galaxy evolution. NGC 5253 and ESO269-G058 are dwarf galaxies in the 
Centaurus group that are presently undergoing episodes of intense star formation, 
and in the current study the old and intermediate age stellar contents of these 
systems are investigated, with the goal of understanding their past evolution 
and placing them in context with other dwarf galaxies.

	Evidence for on-going star formation in NGC 5253 can be traced out 
to $\sim 400$ parsecs from the galaxy center (Calzetti et al. 1997). Still, the 
star-forming activity is highly concentrated in the central regions, and the density 
of ionizing photons suggests that there are 1200 O7 stars in the central pc, and 7000 O 
stars in the inner 20 pc (Turner \& Beck 2004). Nevertheless, while the level of 
star-forming activity in NGC 5253 is impressive by Galactic standards, with the 
possible exception of the nucleus of NGC 5253, there are areas in other galaxies 
that harbour more intense regions of star formation (Alonso-Herrero et al. 2004).

	The nucleus of NGC 5253 consists of a heavily obscured cluster with an age 2 Myr 
and a mass 10$^5$ -- 10$^6$ M$_{\odot}$ (Turner \& Beck 2004; Calzetti et al. 1997; 
Alonso-Herrero et al. 2004). There are other clusters near the center of NGC 5253, 
along with numerous ultracompact HII regions containing 10$^2$ - 
10$^3$ O stars, that have masses approaching 10$^6$ M$_{\odot}$. With the exception 
of the dominant central cluster, which appears to be tidally bound (Harris 
et al. 2004), the smaller clusters may not be long-lived, and Tremonti 
et al. (2001) estimate that they are disrupted on time scales of a few tens of Myr. 

	The bulk of long wavelength emission in NGC 5253 is from 
free-free processes (Turner, Ho, \& Beck 1998), and the absence of 
synchrotron emission ostensibly suggests that star formation has only recently been 
initiated. However, this is almost certainly not the case. The clusters in the central 
regions of NGC 5253 have ages $\leq 20$ Myr (Tremonti et al. 2001; Harris et al. 
2004; Cresci, Vanzi, \& Sauvage 2005), and only $20 - 30\%$ of the lower mass 
clusters have ages $< 7$ Myr (Alonso-Herrero et al. 2004). The oldest star clusters 
have a cumulative mass that is $20\times$ greater than that of the younger clusters, 
and Cresci et al. (2005) argue that the SFR was much higher in the not too distant past. 
The presence of Cepheids in NGC 5253 (Saha et al. 
1995) is evidence of star formation a few tens of Myr in the past, 
while the spectral energy distribution (SED) of the intracluster regions 
within 8 arcsec of the nucleus is indicative of a luminosity-weighted age 
100 -- 500 Myr (Calzetti et al. 1997). Finally, 
NGC 5253 has hosted two Type I SNe in the past $\sim 110$ years, which 
yields a SNe rate that is more than an order of magnitude higher than expected (van den 
Bergh 1980). Such a high Type I SNe rate suggests that NGC 5253 vigorously 
formed stars during intermediate epochs. 

	In contrast to NGC 5253, the young stellar content of ESO269--G058 has not been 
the subject of detailed study. There is a prominent dust lane that slices through the 
center of the galaxy (Sersic 1974; Hawarden et al. 1981), and complicates efforts 
to study the nuclear regions where young stars might be expected. Still, ESO269--G058 
was detected by IRAS at 25, 60, and $100\mu$m (Knapp et al. 1989), and 
these data indicate that there is an elevated SFR in ESO269--G058, although 
not at the level that is seen in NGC 5253. Indeed, 
the total flux from ESO260--G058 in the 12 -- $100\mu$m interval is between 1 -- 2 
orders of magnitude lower than that from NGC 5253. Moreover, the far infrared (FIR) 
SED of ESO269--G058 differs from that of NGC 5253 in the sense that whereas 
the FIR SED of the latter peaks near $60\mu$m, the SED of the former peaks at 
longer wavelengths; thus, dust in ESO260-G058 tends to be at a lower temperature 
than in NGC 5253, suggesting that ESO269--G058 contains a smaller population of hot 
stars.

	The stellar contents of NGC 5253 and ESO269--G058 outside of their 
central regions provide clues into the early evolution of these galaxies, and their 
relationship to other dwarf galaxies. Caldwell \& Phillips (1989), resolved the 
brightest stars in the outer regions of NGC 5253. They detected 
stars with magnitudes consistent with red giants, although a 
more detailed analysis was not possible since these stars are at the 
faint limit of their data. Still, based on the integrated 
colors of the extranuclear region, Caldwell \& Phillips suggest that the outer regions 
of NGC 5253 may contain an intermediate age component, and that NGC 5253 
will evolve into a nucleated dE galaxy.

	Sakai et al. (2004) resolved RGB stars outside of the star-forming 
nucleus of NGC 5253, and used these data to obtain a distance based on the RGB-tip. The 
Sakai et al. (2004) color-magnitude diagram (CMD) shows a well-populated spray of 
stars above the RGB-tip, and a plume of AGB stars that extends to the right of the 
RGB. As for ESO269--G058, the only published study of resolved stars is by Karachentsev 
et al. (2007), who observed the main body of the galaxy with the HST ACS and computed 
a distance modulus using the RGB-tip. Their $(I, V-I)$ CMD shows a broad RGB and a 
plume of bright AGB stars.

	In the present study, deep $r'$ and $i'$ images obtained with 
GMOS on Gemini South are used to probe the properties of the brightest resolved stars in 
the outer regions of NGC 5253 and ESO269-G058. The data are used to characterize the 
properties of stars evolving on (1) the red giant branch (RGB), with the goal 
of probing the metallicity distribution in each galaxy and estimating distances 
using the brightness of the RGB-tip, and (2) the AGB, with the intent of investigating 
the age and spatial distribution of intermediate age stars in each galaxy. 
Comparisons are made with model luminosity function (LFs) to estimate the 
masses of intermediate age components.

	The Centaurus group is viewed at low Galactic latitudes, and so there is 
significant foreground extinction. Throughout this paper it is assumed that 
the foreground extinction towards NGC 5253 is A$_B = 0.24$ while for ESO269--G058 
the assumed extinction is A$_B = 0.46$; both values are from the Schlegel, 
Finkbeiner, \& Davis (1998) maps. As for internal reddening, 
the Balmer decrement of the integrated light near the center of NGC 5253 
suggests that the field population is not heavily reddened; rather, it is only 
the young star clusters that are heavily obscured (Calzetti et al. 1997), and so the 
internal extinction is assumed to be negligible. A similar assumption is made for 
the main body of ESO269--G058, which lies outside of the central dust lane.

	The observations and the procedures used to reduce the data are discussed in 
\S 2, while the CMDs and LFs of stars at various 
galactocentric distances in both galaxies are examined in \S 3. RGB and AGB 
stars are traced out to $\sim 3$ kpc along the minor axis of each galaxy. 
Distances based on the brightness of the RGB-tip are presented in \S 4, while 
a summary and discussion of the results follows in \S 5.

\section{OBSERVATIONS}

	The data were obtained with the Gemini South GMOS as part of program 
GS-2006A-Q-45. The detector in GMOS is a mosaic of three $2048 \times 4068$ EEV CCDs. 
Each raw pixel subtends 0.072 arcsec on a side, and the data were binned $2 \times 
2$ pixel$^2$ during readout to better match the angular sampling with the image quality. 
The GMOS imaging science field covers $5.5 \times 5.5$ arcmin$^2$. 
A complete description of GMOS is given by Crampton et al. (2000).

	The data were recorded through $r'$ and $i'$ filters 
on various nights in March and April 2006. A 5 point dither pattern, which was 
defined to allow the gaps between the CCDs in the detector mosaic to be 
filled during processing, was employed. The total exposure time for each galaxy is 
4050 sec in $r'$ and 4500 sec in $i'$. Stars in the final
NGC 5253 images have a FWHM of 0.8 ($r'$) and 0.7 ($i'$) arcsec, while for 
ESO269--G058 the FWHMs are 0.9 arcsec ($r'$) and 0.8 arcsec ($i'$). The 
NGC 5253 data go slightly fainter than the ESO269-G058 data by virtue of having 
better image quality. The locations of the target fields, which were positioned 
to sample the minor axis of each galaxy, are indicated in Figure 1.

	The data were reduced using standard procedures for CCD imaging. The 
basic steps in the reduction process were (1) the removal of CCD-to-CCD gain 
variations, (2) bias subtraction, (3) the division by flat-field frames, which were 
obtained by observing the twilight sky, and (4) the removal of 
interference fringes from the $i'$ observations. The calibration 
frame used in the last step was constructed by median-combining $i'$ observations of 
various fields, so that stars and galaxies were suppressed while retaining 
interference fringes. The images thus processed were spatially aligned 
and then median-combined. The final images for photometric analysis were obtained by 
trimming the stacked images to the area common to all exposures in the dither pattern.

\section{THE STELLAR CONTENTS OF NGC 5253 AND ESO269--G058}

\subsection{Photometric Measurements}

	The photometric measurements were made with the point spread 
function (PSF) fitting routine ALLSTAR (Stetson \& Harris 1988), 
using PSFs and target lists that were obtained from tasks in 
the DAOPHOT (Stetson 1987) package. The photometric calibration is based 
on Landolt (1992) standard stars, which are observed as part of the baseline calibration 
set for GMOS queue observing. The $r'$ and $i'$ photometric zeropoints 
used in the current dataset have an uncertainity of $\pm 0.03$ magnitudes.

	One goal of this study is to investigate the intrinsic dispersion in the RGB 
sequences of NGC 5253 and ESO269--G058, and this requires that the broadening due to 
random photometric errors be known. This information was obtained by running artificial 
star experiments. The artificial stars, which were assigned brightnesses and colors 
appropriate for stars on the upper RGB in each galaxy, were placed throughout the images 
so that the impact of changing stellar density with distance from the center of 
each galaxy could be assessed. The brightnesses of the artificial stars 
were measured using the same procedures that were employed for actual stars.

\subsection{NGC 5253}

\subsubsection{The CMDs of RGB and AGB Stars}

	The $(i', r'-i')$ CMDs of stars in NGC 5253 are 
shown in Figure 2. The distances quoted in the upper left hand corner of 
each panel are measured along the major axis, and assume (1) an ellipticity 
of 0.57, as measured in the outer regions of the galaxy by Lauberts \& Valentijn 
(1989), and (2) a distance modulus $\mu_0 = 27.5$, which is based on 
the brightness of the RGB-tip (\S 4.1). The spray of objects with $i' 
< 21.5$ in the $0 - 1$ kpc interval are young clusters. 
The effects of crowding on the photometric faint limit are clearly evident 
in the lower portions of the CMDs of the $0 - 1$ and $1 - 2$ kpc intervals.

	The RGB, which is the dominant sequence in the NGC 5253 CMDs when R$_{GC} 
> 3$ kpc, can be traced out to R$_{GC} = 8$ kpc. At larger R$_{GC}$ the CMDs 
are dominated by field stars and background galaxies, with the latter accounting for 
a progressively larger fraction of contaminating objects as one moves towards 
fainter magnitudes. While RGB stars belonging to NGC 5253 may be present 
when R$_{GC} > 8$ kpc, they have a projected density that
is so low that a well-defined RGB sequence is not seen.

	The CMDs of stars in the 3 - 5 kpc intervals in NGC 5253 are compared with 
10 Gyr isochrones from Girardi et al. (2004) in Figure 3. Isochrones with Z = 
0.0001, 0.001, and 0.004 are shown; the solid lines show 
evolution on the first ascent giant branch, while the dashed 
lines shows the portion of the AGB where stellar luminosities 
exceed that at the RGB-tip. The CMD of the 9 -- 10 kpc interval, where there are 
few -- if any -- stars belonging to NGC 5253, is also shown to allow 
contamination from foreground stars and background galaxies to be assessed. The 
locus of RGB stars in the 3 -- 5 kpc interval is reasonably well matched by the Z = 0.001 
([Fe/H] $\sim -1.2$) isochrone, whereas the blue edge of the RGB roughly follows 
the Z = 0.0001 ([Fe/H] $\sim -2.2$) sequence. The uncertainties in the photometric 
measurements predicted from the artificial star experiments are also shown in 
Figure 3, and it is evident that much of the RGB width is due to random uncertainties 
in the photometry.

	The $r'-i'$ distributions of stars with $i' = 24.3 \pm 0.2$ 
(i.e. the upper 0.4 magnitude of the RGB) in three radial intervals are compared 
in Figure 4. To account statistically for sources that do not belong to NGC 5253, the 
color distribution of objects in the $8 - 10$ kpc intervals were scaled to the area 
sampled in each radial interval, and then subtracted 
from the observed distributions. The dotted lines are Gaussians 
that show the distribution expected for an idealized simple stellar system 
(i.e. a system where the RGB has no intrinsic color dispersion) that is 
broadened by photometric errors. The standard deviations of the 
Gaussians were computed from the differences between the known and 
measured brightnesses of artificial stars.

	At $i' = 24.3$ background galaxies are the dominant source of contamination in 
the vicinity of NGC 5253. Star counts from the Robin et al. (2003) model Galaxy
predict that only 2 of the 25 objects with $i' = 24.3 \pm 0.2$ magnitudes 
in the 8 -- 10 kpc field are foreground stars. For comparison, galaxy counts
from Smail et al. (1995) suggest that there should be 30 -- 40 galaxies 
in this brightness range. The dashed lines in Figure 4 show the $r'-i'$ distribution of 
objects in the 8 -- 10 kpc interval, which was used to monitor contamination from 
foreground stars and background galaxies; these relations have been scaled to show the 
fractional contribution that objects in the 8 -- 10 kpc interval make to 
the total number of sources detected in each distance interval. 
It is evident from Figure 4 that objects not belonging to 
NGC 5253 constitute only a modest fraction of the sources detected out to 8 kpc. 

	The mean $r'-i'$ color in the outer regions of NGC 5253 does not change markedly 
with radius, with $<r'-i'> = 0.50 \pm 0.01$. The predicted colors of 10 Gyr isochrones 
with Z = 0.0001 and 0.001 are indicated in Figure 4, and the peak of the color 
distribution is consistent with a metallicity that is slightly higher than Z = 0.001. 
To compute a mean metallicity, a photometric metallicity 
was determined for each star with $i' = 24.3 \pm 0.2$ in the 
$2 - 4$ kpc interval by interpolating between the 10 Gyr Girardi et al. (2002) 
isochrones. The mean of the results was $<[Fe/H]> = -0.95 \pm 0.02$. The 
error is the $1-\sigma$ random uncertainty computed from bootstrap re-sampling, in 
which samples of 500 stars were drawn at random from the stellar catalogue over the course 
of 1000 realizations. It is also worth noting that 
the excellent agreement between the observed color 
distributions and the dotted lines in Figure 4 indicates that the intrinsic 
dispersion in the RGB of NGC 5253 is modest, amounting to only a few hundredths 
of a magnitude in $r'-i'$. This implies that if there are RGB stars with [Fe/H] 
that differ by more than a few tenths of a dex from the mean then they 
have a space density that is substantially lower than that of the dominant component.

	A population of AGB stars is seen above the RGB-tip in the CMDs of the 3 -- 4 
and 4 -- 5 kpc intervals. The majority of these stars fall below the locus of AGB-tip 
values for a 10 Gyr population, indicating that they belong to an old population. 
Still, a scattering of stars is seen above the locus of 10 Gyr AGB-tip 
values in the 3 -- 4 and 4 -- 5 kpc intervals. These stars 
occur in numbers that exceed what is seen in the 9 -- 10 kpc 
field at comparable brightnesses, suggesting that a population of stars with 
an age $< 10$ Gyr is present at intermediate distances from the center of NGC 5253.
The $i'$ brightness of the AGB-tip in these intervals is 
consistent with an age of a few hundred Myr (\S 3.2.2). 

	The AGB stars in the 3 -- 4 and 4 -- 5 kpc intervals 
span a wide range of $r'-i'$ colors, and the isochrones suggest that these 
stars have a larger range of metallicities than predicted from the RGB. This being said, 
caution should be excercised when interpreting the colors of upper AGB stars, 
as many are photometric variables, and their 
colors will change as they cycle through their light curves. 
Hence, they are not ideal tracers of metallicity.

	The metallicity computed from the color distributions in Figure 4 assumes that 
the RGB is made up of stars that have an old age, whereas the presence of bright AGB 
stars indicates that an intermediate age component is present. If the intermediate age 
component is sufficiently large then it will skew the metallicity estimate. The impact of 
adopting a markedly lower age for the main body of RGB stars is demonstrated in Figure 4, 
where the dashed arrow indicates the color expected for a Z=0.001 RGB 
with an age of 2 Gyr. If the mean age of NGC 5253 is as low as this -- which would make 
NGC 5253 a truely remarkable galaxy -- then the metallicity computed by assuming an 
old age will be underestimated by $\sim 0.6$ dex.

	The metallicity inferred for the RGB of NGC 5253 is likely not greatly skewed by 
the presence of intermediate age stars, as these objects constitute only a minor 
portion of the stellar mass in NGC 5253. This claim is 
substantiated in \S 3.2.2, where the size of the intermediate 
age component is estimated from comparisons with model LFs.
Hints that the intermediate age population does not dominate the 
stellar content can also be obtained from the CMDs. Rejkuba et al. 
(2006) used the ratio of bright AGB stars to stars on the 
upper RGB, which they refer to as P$_{IA}$, to gauge the contribution 
that intermediate age stars make to two Centaurus group dEs. Rejkuba et al. (2006) 
find that P$_{IA} \sim 0.05 - 0.10$ in these objects, and conclude that 
the fraction of intermediate-age stars is likely $\sim 15\%$ based on models of 
stellar evolution. In the outer regions of NGC 5253, 
which is the portion of the galaxy that is furthest 
from the region of present-day star formation, the GMOS data indicate that P$_{IA} 
\sim 0.04$, which is smaller than in the dEs studied by Rejkuba et al. (2006). 
Therefore, the intermediate age component in the outer regions of NGC 5253 
is likely smaller than in the Rejkuba et al. (2006) galaxies. 

\subsubsection{The Number Density of Red Giants}

	The $i'$ LFs of objects in NGC 5253 that have $r'-i'$ between -0.2 and 1.2 are 
shown in Figure 5; the LFs of the 0 -- 1 and 1 -- 2 kpc intervals are not considered, 
as crowding hinders the ability to sample RGB stars in these regions. Background 
and foreground objects have been removed in a statistical fashion by scaling 
the LF of the 8 - 10 kpc interval to match the area sampled in each annulus and then 
subtracting the results from the observed LFs. The dashed lines show a power-law 
that was fit to the number counts of the mean LF of objects 
with R$_{GC}$ between 4 and 8 kpc in the $i'$ = 24.5, 25.0, and 25.5 bins. 
The y-intercept was then shifted to match the number counts in the 
$i'$ = 24.5, 25.0, and 25.5 bins in each R$_{GC}$ interval.

	The drop in stellar density associated with the RGB discontinuity is clearly 
seen in Figure 5, where the LFs fall below the dashed lines when $i' \leq 24.0$. 
The magnitude intervals above the RGB-tip are populated by AGB stars 
(Figure 3), which evolve at a faster pace than first ascent giants, and so 
have a lower space density. The dotted line shows the power-law 
computed from the RGB LF, but with the y-intercept shifted 
to account for a population of objects with a space density that is one fifth 
that of RGB stars in each distance interval; such a difference in space density 
more-or-less matches AGB number counts in old systems, and is roughly consistent with 
stellar evolution models (e.g. Davidge et al. 2004; Davidge 2006). 
It should be emphasized that the dotted line is shown simply as 
a point of reference, and that the ratio of AGB to RGB stars will depend on the 
star-forming history of a system. Indeed, if an intermediate age population 
is present that accounts for more than a few percent of the total galaxy mass then 
the observed LF will likely fall above the dotted line (see Appendix). 

	Preliminary ages can be deduced from the brightness of the AGB-tip, and predicted 
AGB-tip brightnesses from the Girardi et al. (2004) isochrones for various values of 
log(t$_{yr}$) are indicated at the top of Figure 5. The age scale is from models with 
Z = 0.001, although the brightness of the AGB-tip is not sensitive to metallicity 
when [Fe/H] $< -1$. The brightest AGB stars in the 3--4 and 4--5 kpc intervals appear 
to have log(t$_{yr}) \sim$ 8.2, while in the $5-6$ kpc interval the AGB-tip 
brightness is suggestive of log(t$_{yr}) \sim$ 9. These ages are likely lower limits 
(see below). There is no statistically robust 
evidence for an intermediate age population at larger distances, so that if an 
intermediate age population is present in the 6 -- 8 kpc interval then it must 
have a space density such that the brightest AGB stars do not 
stand out statistically against foreground and background objects. 

	A caveat when considering ages estimated from the AGB-tip brightness is that 
many of the most evolved AGB stars are photometric variables. This variability biases 
age estimates to younger values, as variables near the 
peak of their light curves will have brightnesses that are comparable to those of 
younger, non-variable objects. In addition to variability driven by ionization effects, 
AGB stars also show photometric variations due to thermal pulses. However, the photometric 
signatures of these events are very short-lived, affecting the luminosity only 
during $\sim 10^{-3}$ of the time between pulses (e.g. Boothroyd \& Sackmann 1988). Hence, 
thermal pulses likely do not significantly bias AGB-tip brightness measurements 
in external galaxies. 

	The impact of variability on ages measured from the brightness of the 
AGB-tip can be estimated from photometric observations of LPVs. Reid, Hughes, \& Glass 
(1995) summarize photometry of LPVs in the northern regions of the LMC, and the 
entries in their Table 1 indicate that these stars have typical rms dispersions 
of $\pm 0.3$ magnitudes about their mean $I$ magnitudes. Given that $\Delta(I) \sim 
\Delta(i')$, then a comparison with the age scale at the top of Figure 5 indicates 
that photometric variability may skew age estimates made from the AGB-tip to younger 
values by a few tenths of a dex.

	The stars with $i' < 23$ and R$_{GC}$ between 3 and 6 kpc have number densities 
that place them 0.5 -- 1.0 dex below the dotted line in Figure 5. Moreover, when R$_{GC} 
< 6$ kpc the AGB component does not follow a single power-law; rather, at the 
bright end the LF is more-or-less flat, but then steepens in the 1 magnitude interval 
above the RGB-tip, where the LFs intercept the dashed line. This change in the LF of AGB 
stars suggests that there are {\it at least} two AGB components with very different ages 
in NGC 5253. Based on the models described in the Appendix, the 
majority of AGB stars in the one magnitude interval immediately above 
the RGB-tip likely belong to the dominant old component, while those with $i' < 23$ 
originate from a younger population (or younger populations).

	The fraction of the total mass in any radial interval that belongs 
to the intermediate age stars can be estimated by comparing the 
observed LFs with models. In the Appendix, model LFs are 
constructed by combining the LF of an intermediate age population that has 
an age $\sim 0.7$ Gyr with the LF of an old population, and the results are shown in 
Figure A1. Comparisons with Figure A1 indicate that the number density of bright 
AGB stars in NGC 5253 results from an intermediate age component 
that contributes only a few percent of the total stellar mass in the $3 - 6$ kpc 
interval. That the fractional contribution from intermediate age stars is small 
also indicates that the metallicity derived from the color of the RGB in \S 3.2.1 is 
likely not skewed by the age-metallicity degeneracy for RGB colors.

	Based on integrated colors obtained from aperture photometry, Caldwell \& 
Phillips (1989) argue that an intermediate age population is present in the outer regions 
of NGC 5253. While studies of resolved stars provide the means of testing this 
hypothesis, the use of relatively rare AGB stars as probes of stellar content in the 
present study limit the ability to draw firm conclusions about the spatial distribution 
of an intermediate age population to that portion of NGC 5253 with R$_{GC} < 6$ kpc. At 
larger R$_{GC}$ the use of AGB stars as probes of stellar content becomes problematic. 
Indeed, if a population of bright AGB stars like those at smaller R$_{GC}$ were 
uniformly mixed with the RGB component then they would not produce a detectable 
number of objects when R$_{GC} \geq 6$ kpc.

\subsection{ESO269--G058}

\subsubsection{The CMDs of RGB and AGB Stars}

	The $(i', r'-i')$ CMDs of stars in ESO269--G058 are shown in Figure 6. 
The distance interval listed in each panel assumes an ellipticity of 0.29 
and a distance modulus $\mu_0 = 27.9$; the former was 
measured from the red DSS image of ESO269--G058, whereas the latter is based on 
the brightness of the RGB-tip (see below). With a Galactic latitude near $16^o$, 
ESO269--G058 is much closer to the Galactic plane than NGC 5253, and so there is 
a greater amount of foreground star contamination in the ESO269--G058 CMDs. 
The impact of foreground star and background galaxy contamination is smallest in the 
CMDs that sample the innermost regions of the galaxy, as smaller angular areas 
are covered than in the outer regions of the galaxy.

	A number of blue objects with $r'-i' \sim -0.3$, which 
are probably compact young star clusters, are seen in the $0 - 1$ kpc CMD. 
The RGB is the dominant sequence in the ESO269--G058 CMDs when R$_{GC} > 2$ 
kpc, and RGB stars are seen out to R$_{GC} \sim 6$ kpc. The 
CMDs of stars in the 2 - 4 kpc intervals are compared with 10 Gyr isochrones from Girardi 
et al. (2004) in Figure 7. As in Figure 3, isochrones with Z = 0.0001, 0.001, and 0.004 
are plotted; the solid lines show evolution on the first 
ascent giant branch, while the dashed lines show evolution on the 
upper AGB. The CMD of objects in the 8 -- 9 kpc interval is also shown in 
Figure 7 to allow the reader to assess contamination from foreground stars. 
The three R$_{GC}$ intervals used to construct Figure 7 sample similar areas on the sky. 

	The RGB in ESO269--G058 is broader than in NGC 5253, and the RGB 
locus falls between the Z = 0.001 ([Fe/H] $\sim -1.2$) and Z = 0.0001 
([Fe/H] $\sim -2.2$) isochrones. There is a population of AGB stars above the RGB-tip in 
the 2 --3 kpc interval, and the peak brightness is consistent with an age $< 
10$ Gyr. Thus, ESO269--G058 likely harbours an intermediate age population. 

	The $r'-i'$ distributions of stars in ESO269--G058 with $i' = 24.9 \pm 0.2$ 
in two radial intervals are compared in Figure 8. As in Figure 4, the color distribution 
of objects in the outermost portions of ESO269--G058 were scaled to match the area 
sampled by the two annuli with smaller R$_{GC}$, and these were then subtracted from the 
observed distributions to suppress contamination from foreground stars and background 
galaxies. The dotted lines in Figure 8 are Gaussians that show the distributions 
expected for an idealized simple stellar system that is broadened 
by photometric errors. If the observed distributions are broader than the dotted 
lines then it would indicate that a significant intrinsic dispersion in $r'-i'$ is 
present among stars in ESO269--G058. 

	The dashed lines in Figure 8 show the color distributions of sources in 
the 7 - 9 kpc interval, where the number of sources belonging to ESO260--G058 
is modest. As in Figure 4, these distributions have been scaled to show the fractional 
contribution made by contaminating objects in each distance interval. The 
Robin et al. (2003) star count models and Smail et al. (1995) galaxy counts 
predict that galaxies account for $\sim 90\%$ of the sources with 
$i' = 24.9 \pm 0.2$ magnitude in the vicinity of ESO269--G058. 

	Figure 8 confirms the impression from the CMDs that the 
mean $r'-i'$ color in the outer regions of ESO269--G058 is consistent with 
a metallicity that is between Z = 0.001 and Z = 0.0001.
The mean metallicity of stars in the 2 -- 4 kpc interval 
was computed using the procedure described in \S 3.2.1, and the result is 
$<[Fe/H]> = -1.77 \pm 0.05$; as in \S 3.2.1 the uncertainty was computed by applying 
a bootstrap procedure. It is also apparent from Figure 8 that the width of the 
RGB in ESO269--G058 is dominated by photometric uncertainties; therefore, as in 
NGC 5253 the density of stars with [Fe/H] that differ from the mean 
value by more than a few tenths of a dex is small in ESO269--G058.

	As with NGC 5253, the age-metallicity degeneracy introduces uncertainties into 
the metallicity estimated for ESO269-G058 from the mean color of the RGB. As was done 
for NGC 5253, P$_{IA}$ was computed in the outer regions of ESO269-G058, and the result 
is comparable to that in the outer regions of NGC 5253. Therefore, 
intermediate age stars likely are not large contributors 
to the color distribution in the lower panel of Figure 8. Still, the metallicity 
should be considered to be a lower limit given that some intermediate age stars are 
present.

\subsubsection{The Number Density of Red Giants}

	The $i'$ LFs of objects with $r'-i'$ between -0.2 and 1.2 in ESO269 -- G058 are 
shown in Figure 9. Contamination from background and foreground objects has been 
removed by subtracting the LFs of sources with R$_{GC}$ between 7 and 9 kpc. 
The ESO269--G058 data do not sample as much of the RGB as was covered with the 
NGC 5253 data, due to the greater distance of ESO269--G058, coupled with the 
slightly poorer image quality of the ESO269--G058 data. 
Rather than attempt to fit a power-law to the RGB portion of the ESO269--G058 
LF, the dashed lines in Figure 9 are power-laws 
with the exponent measured from the NGC 5253 data, but with the y-intercept 
shifted to match the number counts in the $i'$ = 25.0 bin in each distance interval. 

	The RGB-tip produces a discontinuity near $i' = 24.5$ in the 
ESO269--G058 LFs, and the number density of AGB stars in the $i' = 24$ bin 
is consistent with that expected from a population of old AGB stars. There is also 
evidence of an intermediate age population, in the form of objects with $i' < 23.5$. 
Based on the peak brightness of the AGB sequence, it is likely that the
dominant intermediate age component is not as young as in 
NGC 5253. In fact, the peak AGB brightness suggests that the youngest stars in 
the 3 -- 4 kpc interval in ESO269--G058 have an age in excess of 1 Gyr. A comparison 
with the model LFs discussed in the Appendix suggest that -- as in NGC 5253 -- the 
intermediate age component contributes only a few percent of the total stellar mass.

\section{DISTANCE MODULI}

\subsection{NGC 5253}

	The distance to NGC 5253 can be estimated from the brightness of the RGB-tip, 
which can be measured by applying an edge-detection filter to LFs (e.g. Lee, 
Freedman, \& Madore 1993a). The result of convolving the $i'$ LF of stars 
in NGC 5253 that have R$_{GC}$ between 4 and 6 kpc with a three point Sobel kernel 
is shown in Figure 10. The convolved signal shows a distinct peak that spans 
the $i' = 24.0$ and $i'=24.2$ bins, and a weighted mean of the 
signal in these bins indicates that the RGB-tip occurs at $i' = 24.09 \pm 0.10$. 
The error quoted here includes the uncertainty in measuring the RGB-tip brightness 
from the filter-convolved LF, which is taken to be the half width of the 
peak in the edge-detection signal (Lee et al. 1993a), 
combined with the uncertainty in the photometric calibration. 

	The standard empirical calibration of the  RGB-tip brightness is in the 
Kron-Cousins $I$ filter, and so the $i'$ RGB-tip measurement of NGC 5253 
has been transformed into this system. Given that stars 
near the RGB-tip in NGC 5253 have $(r'-i')_0 \sim 
0.46$ (e.g. Figure 4), then the Fukugita et al. (1996) transformation equations, coupled 
with the colors of solar neighborhood giants from Bessell \& Brett (1988), indicate that 
the measured RGB-tip brightness $i' = 24.09$ corresponds to $I = 23.59$. The RGB-tip 
occurs at M$_I = -4 \pm 0.1$ when [Fe/H] $\leq -0.7$ (Lee et al. 1993a). Therefore, 
after accounting for foreground extinction, the RGB-tip distance modulus of 
NGC 5253 is $27.48 \pm 0.14$. No correction has been made for internal extinction, 
and it should be noted that the dust in NGC 5253 appears to be concentrated 
near the youngest nuclear clusters (Calzetti et al. 1997); hence, there is 
unlikely to be significant dust obscuration in the outer regions of the galaxy.

	The RGB-tip distance modulus computed from the GMOS data differs 
from that measured by Sakai et al. (2004). Indeed, the GMOS data indicate that the 
RGB-tip in NGC 5253 occurs at I = 23.6, whereas Sakai et al. (2004) conclude that the 
RGB-tip occurs near I = 24. This 0.4 magnitude difference 
is a result of different features being identified as the RGB-tip. In the 
current study, the RGB-tip is placed at the base of the 
RGB discontinuity in the LF, which is the usual convention 
(e.g. Figure 2 of Lee et al. 1993a; Figures 1 -- 6 of McConnachie et al. 2005). 
The logarithmic LF of NGC 5253 in Figure 6 of Sakai et al. (2004) shows a 
clear inflexion point at $I = 23.6$ at the base of the RGB LF. An inspection of the LFs 
and the associated plots of edge-detection statistics used by Sakai et al. 
(2004) indicates that their technique also usually triggers on the base of the RGB. 
However, in their Figure 6 it is evident that the RGB-tip in NGC 5253 was placed at the 
top of the RGB LF discontinuity; had Sakai et al. (2004) located the RGB-tip at 
the base of the RGB discontinuity then their distance modulus 
would agree with that found here.

	A modest number of Cepheids have been discovered in NGC 5253 (Saha et al. 1995), 
and the distance modulus computed from these objects can be compared with the RGB-tip 
distance modulus. Unfortunately, Saha et al. (2006) show in their Table 4 that the 
Cepheid distance modulus can change by almost 0.5 magnitudes, depending on the sample of 
objects selected and the filter used. In fact, the Cepheid-based distance estimates 
advocated by various groups cluster near either the GMOS or Sakai et al. (2004) RGB-tip 
distance moduli. For example, Saha et al. (2006) find that 
$\mu_0 = 27.61$ if all 14 Cepheids from Saha et al. (1995) are used with an 
LMC-based calibration, but that $\mu_0 = 28.09$ if a subset of five objects is 
retained; the latter value is prefered by Saha et al. (2006). 
For comparison, in the summary paper of the Key Project distance scale, 
Freedman et al. (2001) compute a final distance modulus of 27.56 for NGC 5253 
from a subset of 4 Cepheids.

	The $i'$ magnitude of the RGB-tip was transformed into the Kron-Cousins I--band 
to permit an accepted empirical calibration to be used to compute the distance modulus, 
and this introduces uncertainties into the distance estimate. The need for such a 
transformation could be avoided if an empirical calibration for the RGB-tip brightness 
in $i'$ was employed; however, there is as yet no such calibration known to us. Still, 
a distance modulus can be computed from stellar evolution models. The 
isochrones discussed by Girardi et al. (2004) predict that M$_{i'}^{RGBT} \sim -3.47$ for 
old stellar systems with [Fe/H] $\sim -1.2$. The distance modulus 
computed with this calibration is $\mu_0 = 27.45$. This 
agrees with the distance modulus computed from the $I-$band calibration, suggesting 
that the transformation from $i'$ to $I_{KC}$ probably does not introduce uncertainties 
larger than a few hundredths of a magnitude into the distance modulus measurement.

\subsection{ESO269--G058}

	The $i'$ LF of stars with R$_{GC}$ between 4 
and 5 kpc in ESO269--G058 is shown in the top panel of 
Figure 11, while the result of convolving the LF with a 3 point Sobel kernel is shown in 
the bottom panel. The peak of the convolved signal spans 
the $i' = 24.4$, 24.6, and 24.8 bins, and a weighted mean of the signal in these bins 
indicates that the RGB-tip occurs at $i' = 24.64 \pm 0.15$. As with NGC 5253, the error 
includes the uncertainty in the measurement of the RGB-tip from the edge-convolved 
signal, and the error in the photometric calibration. The transformation procedure 
described in \S 4.1 indicates that the RGB-tip brightness corresponds to I = 24.14, 
and the RGB-tip distance modulus of ESO269--G058 is then $27.93 \pm 0.18$ 
based on the Lee et al. (1993a) calibration. As with NGC 
5253, no correction has been made for internal extinction. This distance modulus 
is in excellent agreement with that found by Karachentsev et al. (2007).

\section{SUMMARY \& DISCUSSION}

	Deep $r'$ and $i'$ images are used to probe the stellar contents in the outer 
regions of the Centaurus group dwarf galaxies NGC 5253 and ESO269--G058. First ascent 
giants are detected out to distances of 6 -- 8 kpc along the major axes of both 
galaxies, and the mean metallicities and intrinsic metallicity dispersions of RGB stars 
in the outer regions of these galaxies are found to be similar. Distances have been 
computed using the brightness of the RGB-tip. While the distance modulus computed for 
ESO269-G058 is consistent with a previously published  RGB-tip distance estimate, the 
RGB-tip distance modulus computed for NGC 5253 is not consistent with that measured by 
Sakai et al. (2004). The properties of the brightest AGB stars have 
also been investigated, and evidence is found for a spatially extended intermediate age 
population, which accounts for a few percent of the total stellar mass in the outer 
regions of both galaxies. Various aspects of the evolution of NGC 5253 and ESO269--G058 
are discussed in the following sub-sections.

\subsection{The Recent Star Forming Histories of NGC 5253 and ESO269--G058}

	The current episode of star formation in NGC 5253 has not been an isolated event. 
Indeed, studies of the clusters in the nuclear regions of NGC 
5253 reveal that star formation occured over a period spanning at least a few tens of 
millions of years (e.g. Tremonti et al. 2001; Harris et al. 2004; Cresci et al. 2005). 
The GMOS data presented here indicate that large scale star forming activity in NGC 5253 
likely occured over an even more extended period of time. In particular, there are 
a significant number of AGB stars with M$_{i'} \sim -6.5 \pm 0.25$ 
with R$_{GC}$ between 3 and 5 kpc, and comparisons with model LFs suggest that 
these are the brightest members of a star forming episode that may have increased 
the stellar mass by a few percent.

	The Girardi et al. (2004) isochrones 
suggest that the brightest AGB stars in NGC 5253 belong to a population with an 
age $\sim 0.2$ Gyr. It is curious that no clusters with this age have been discovered in 
NGC 5253 (Harris et al. 2004), although perhaps these objects have been disrupted (e.g. 
Tremonti et al. 2001). The elevated level of SN Ia activity in NGC 5253 is also 
evidence for enhanced star forming activity during intermediate epochs. Age dating 
of SN Ia is very uncertain, and SN Ia activity may peak $2 - 4$ Gyr after their 
progenitors form (Strolger et al. 2004), although the timing depends on the star 
formation history (Matteucci \& Recchi 2001). Given these uncertainties, we 
speculate that the progenitors of the two recent SN Ia in NGC 5253 may have formed 
at the same time as the progenitors of the brightest AGB stars. 

	Elevated SFRs like those in NGC 5253 and ESO269-G058 could be 
triggered by tidal interactions, and van den Bergh (1980) argues that NGC 5253 
likely interacted with M83. However, the dynamical properties of HI in NGC 5253 
are also consistent with the accretion of a gas-rich companion (Kobulnicky \& 
Skillman 1995). The cause of the high SFR in NGC 5253 notwithstanding, 
it is perhaps not surprising that star formation has 
occured continually or episodically over the past few hundred Myr in 
NGC 5253, as similar star formation histories are 
seen in other active star-forming dwarf galaxies. Indeed, 
elevated star formation levels have been traced back hundreds of Myr 
in M82 and NGC 3077 (Parmentier, de Grijs, \& Gilmore 2003; Harris et al. 2004), 
which is when these galaxies are thought to have first interacted with M81 (e.g. 
Brouillet et al. 1991; Yun, Ho, \& Lo 1994). In fact, NGC 5253 and NGC 
3077, which is roughly at the same distance as NGC 5253, have similar 60 and 100$\mu$m 
fluxes (Sanders et al. 2003). The density of intermediate age stars in 
ESO269-G058 suggests that this galaxy may have experienced a SFR like that in 
NGC 5253 roughly $\sim 1$ Gyr in the past. While NGC 5253 and ESO 269--G058 may not 
currently be in close proximity to much larger galaxies, the elevated levels of star 
formation may have had their origins hundreds of millions of years or more in the past.

	The GMOS data indicate that bright AGB stars, and hence an intermediate-age 
population, is present well outside of the nuclear regions of NGC 5253 and ESO269--G058.
There are indications that the presence of intermediate age stars at large distances 
from the centers of dwarf galaxies may be a common phenomenon. Consider the dE companions 
of M31, where C stars have been traced out to linear distances in excess of 
2 kpc in NGC 185 and NGC 205 (Battinelli \& Demers 2004a; Demers, Battinelli, \& 
Letarte 2003), and out to linear distances in excess of 4 kpc in NGC 147 (Battinelli \& 
Demers 2004b). Thus, an intermediate age component that extends over spatial scales of a 
few kpc is not uncommon in dEs.

	The intermediate age stars that are seen in the outer regions of the Centaurus 
and M31 dwarfs need not have formed {\it in situ}. The young stars that form near the 
centers of galaxies will drift away from their places of birth as their orbits are heated 
by interactions with molecular clouds; indeed, Tremonti et al. (2001) argue that the 
conditions near the center of NGC 5253 are such that even massive, compact clusters 
are disrupted. Caldwell \& Phillips (1989) measure 
the stellar velocity dispersion in NGC 5253 to be $46 \pm 5$ km sec$^{-1}$. 
If intermediate age stars in NGC 5253 attain even a fraction of this motion in the 
form of a radial component then over the course of a few hundred million years 
they will be re-distributed throughout the galaxy.

\subsection{NGC 5253 and ESO269--G058 as Dwarf Ellipticals}

	The morphological properties of NGC 5253 and ESO269--G058 
are suggestive of a relationship to dEs, and 
Caldwell \& Phillips (1989) argue that NGC 5253 is a dE that is currently 
experiencing an episode of intense star formation. The defining 
characteristic of elliptical galaxies is a smooth light profile that is devoid of 
discrete structure (e.g. Hubble 1926). Recent star formation in NGC 5253 and ESO269--G058 
is restricted to the inner regions of these galaxies, and 
this is where departures from a smooth light profile are 
seen. While an intermediate age population is seen in the outer regions of these 
galaxies, there are no discrete clumps of star formation, and the outer 
isophotes are smooth.

	NGC 5253 and ESO269--G058 have markedly higher levels of nuclear star formation 
than in Local Group dEs. Indeed, after adjusting for differences in distance, 
the measurements listed in Knapp et al. (1989) indicate that 
the flux from NGC 5253 in the $25 - 100\mu$m wavelength 
region is 2 - 3 orders of magnitude higher than that from 
NGC 205, while the FIR flux from ESO269--G058 is roughly an order of magnitude higher 
than that from NGC 205. Still, the impact of recent star formation on galaxy morphology 
is suppressed in the near-infrared, where older populations make a larger contribution 
to the integrated light than at visible wavelengths, and the 2MASS images of NGC 5253 
and ESO269--G058 are very similar to those of the Local Group dEs NGC 185 and NGC 147. 

	The near-infrared brightnesses of NGC 5253 and ESO269--G058 are also similar 
to those of Local Group dEs. Adopting a distance modulus of 27.5 then 
M$_K \sim -19.3$ for NGC 5253 (Jarrett et al. 2003), while M$_K = -17.3$ for 
ESO269--G058 (Jarrett et al. 2000) if $\mu_0 = 27.9$. These brightnesses are 
affected by recent star formation, and should be `faded' if 
comparisons are to be made with the M31 dE's, which have 
comparatively modest levels of nuclear star formation. 
Two limiting cases, in which recently formed stars account for either 1\% or 
10\% of the total stellar mass, are considered to explore the effects of evolutionary 
fading; these contributions from young stars bracket 
plausible values for NGC 5253, which will have the largest 
fading correction. Models discussed by Leitherer et al. (1999) suggest that if the young 
stars in NGC 5253 have an age of $\sim 10$ Myr and make up just 1\% of the stellar 
mass then M$_K$ will fade by $\sim 0.4$ magnitudes over 1 Gyr once star formation is 
terminated. For comparison, if young stars make up 10\% of the total mass then M$_K$ will 
fade by $\sim 1.7$ magnitude. Adopting these as limits for NGC 5253, then the faded M$_K$ 
of this galaxy will be in the range --17.6 to --18.9. The effects of fading will be much 
less dramatic for ESO269--G058, as the FIR flux suggests that it has a much lower 
SFR (see above), and fading corrections will likely be no larger than a few 
tenths of a magnitude. The integrated brightnesses of the M31 dEs range between 
M$_K = -17.3$ (NGC 147) and M$_K = -18.9$ (NGC 205), and so overlap with the range of 
plausible faded brightnesses of NGC 5253 and ESO269--G058.

	Davidge (2004) points out that the globular cluster content of amorphous 
star-forming dwarf galaxies may provide clues into their past histories, and a search 
for classical globular clusters near NGC 5253 and ESO269--G058 is one means of testing 
the hypothesis that these galaxies are dEs. The 
dE companions of M31 have globular cluster specific frequencies S$_N \sim 
2 - 6$, whereas late-type disk systems tend to have S$_N < 1$ (e.g. Harris 1991). 
If NGC 5253 and ESO269--G058 are classical dEs that are experiencing 
elevated lavels of star formation then they should each have $4 - 9$ old 
globular clusters. This is a lower limit, as it might be anticipated that 
dEs that experienced especially vigorous levels of star 
formation in the past might have a very high S$_N$ if they formed 
long-lived compact star clusters, like the central cluster in NGC 5253.

	We close this section by noting that NGC 5253 and ESO 269--G058 are 
located at moderately large distances from the dominant members of the Centaurus group. 
More specifically, the entries in Table 2 of Karachentsev et al. 
(2007) place NGC 5253 at a distance of 0.7 Mpc from Cen A and 1.6 Mpc from M83. 
As for ESO269--G058, this galaxy is 0.3 Mpc from Cen A, and 1.6 Mpc from M83. The 
locations of both galaxies suggest that they have spent most of their lives free of 
interactions with these large galaxies, and the narrow metallicity dispersions 
of NGC 5253 and ESO269--G058 are consistent with such a picture.
Models of the formation and early evolution of isolated 
spheroidal galaxies predict that isolated systems experience a rapid early 
chemical enrichment (e.g. Chiosi \& Carraro 2002), but only low levels of subsequent 
star formation, that became more centrally concentrated with time (Hensler, 
Theis, \& Gallagher 2004). Thus, if dE galaxies evolve in isolation for a significant 
fraction of the age of the Universe then the RGB stars in their outer regions will have 
narrow metallicity distributions that are a consequence of their rapid early enrichment 
and subsequent subdued levels of star formation.

	There have only been limited efforts to simulate the metallicity distribution 
functions (MDFs) of isolated spheroidal systems. Font et al. (2006) investigate 
the chemical evolution of present-day Galactic satellites. The models constructed 
by Font et al. (2006) are of interest in the context of isolated systems since 
they predict that the present-day satellites have only recently been accreted by the 
Milky-Way, and so have evolved largely in isolation. Star formation in these models 
is truncated when the satellites are accreted by the Milky-Way, and so the 
MDFs of the surviving satellites in these simulations is that which was imprinted 
early-on. The MDFs of the `surviving' Milky-Way satellites, which 
are shown in Figure 6 of Font et al. (2006), are those of stars in all of 
the surviving satellites. However, in most of the simulations the satellite 
population is dominated by one or two massive survivors, and so the MDFs are 
essentially those of these massive systems. The dominant survivors in these 
simulations have masses comparable to those of NGC 5253 and ESO269--G058, 

	The Font et al. (2006) simulations predict that systems that spend most of their 
times in isolation will have metallicity dispersions in the range $\pm 0.2 - 0.3$ 
dex. Not only is this consistent with the metallicity dispersions estimated for 
NGC 5253 and ESO269--G058 in \S 3, but it is also consistent with what is seen 
in at least some of the dE companions of M31. In particular, 
the metallicity dispersion in NGC 147 and NGC 185 is $\pm 0.3$ dex 
(Mould, Kristian, \& Da Costa 1983; Lee, Freedman, \& Madore 1993b; Davidge 1994).
For comparison, NGC 205, which is the M31 dE that has likely had the highest degree of 
interaction with M31, has a metallicity dispersion that is at least $\pm 0.5$ dex 
(Mould, Kristian, \& Da Costa 1984). 

\acknowledgments{Sincere thanks are extended to the anonymous referee, whose 
critical comments greatly improved the paper.}

\appendix

\section{Modelling the Luminosity Function of an Old $+$ Intermediate Age System}

	NGC 5253 and ESO269--G058 contain AGB stars at intermediate galactocentric 
distances that belong to an intermediate age component. It is of obvious interest 
to estimate the fraction of the total stellar mass 
in each galaxy that is made up by intermediate age stars, not only to 
probe the level of any past star-forming activity, but also because the presence of 
large numbers of intermediate age stars may skew efforts to estimate metallicity (\S 3). 
To allow constraints to be placed on the size of such a population, the 
$i'$ LF of a simple two-component system that consists of an intermediate age and an old 
stellar population has been investigated.

	Ferraro et al. (2004) discuss near-infrared photometry of six intermediate age 
clusters in the LMC, and their data serve as the basis for constructing the 
intermediate age LF that will be used in the models. The clusters studied by Ferraro et 
al. (2004) have known integrated brightnesses and ages, and the photometry has a high 
level of completeness. The s-parameters (Elson \& Fall 1985) of these clusters 
suggest that four of them have ages $\sim 0.6 - 0.7$ Gyr, while the total age range for 
all six clusters is between 0.5 and 0.9 Gyr. Adopting cluster $K$ magnitudes from Pessev 
et al. (2006) and Persson et al. (1983), then the composite LF of all six clusters 
should represent that of a system with a luminosity-weighted age $\sim 0.7$ Gyr, 
based on the ages given in the fourth column of Table 4 from Ferraro et al. (2004).

	The $K$ brightnesses and $J-K$ colors tabulated by Ferraro et al. (2004) were 
used to compute an $i'$ magnitude for each object with $K_0 < 13.5$ that is 
located within the AGB and RGB boundaries indicated in Figure 5 of Ferraro et al. (2004). 
This transformation was done using colors for giant stars from Bessell \& Brett (1988) 
and Cox (2000), combined with the relevant equations from Fukugita et al. 
(1996). These calibrations hold for stars with $J-K < 1.3$. However, a handful of 
AGB stars in the LMC clusters have $J-K < 1.3$. Transformed $i'$ magnitudes 
for these stars were computed using relations derived from the simulataneous photometric 
measurements of SMC and LMC long period variables tabulated 
by Wood, Bessell, \& Fox (1983).

	The LF of the old population is based on the number counts of stars in the 
metal-poor globular cluster NGC 2419, which were discussed by Davidge et 
al. (2002). While this cluster is more metal-poor than the LMC 
clusters studied by Ferraro et al. (2004), the number density of giants does not vary 
greatly with metallicity. The relative masses of the intermediate age and old systems 
were computed using $K-$band M/L ratios from Maraston (2005), and the results 
of combining the old and intermediate age LFs with various mass mixtures are 
shown in Figure A1. The number counts in this figure refer to those expected 
per 0.5 M$_{i'}$ magnitude interval in a system with a total mass that is approximately 
10$^5$ M$_{\odot}$.

	It can be seen from Figure A1 that an intermediate age population that 
accounts for only 1\% of the total stellar mass can produce a bright AGB sequence 
that falls $\sim 1$ dex below the dotted lines in Figures 5 
and 9. Thus, the intermediate age populations in NGC 5253 and ESO269--G058 account for 
only a few percent of the total stellar mass. Moreover, intermediate age 
populations need not be the dominant component in a stellar system in 
order to have a significant impact on the LF at intermediate brightnesses. 
If an intermediate age population accounts for only 10\% of the stellar mass then the 
model LF becomes very smooth; the discontinuity due to the RGB-tip largely disappears and 
the LF falls well above the dotted line in the brightness interval 
immediately above the RGB-tip. The presence of a prominent RGB-tip is thus also 
an indicator that stars with ages $< 1$ Gyr make only a modest contribution to the 
total system mass.

\clearpage
\begin{figure}
\figurenum{1}
\epsscale{0.75}
\plotone{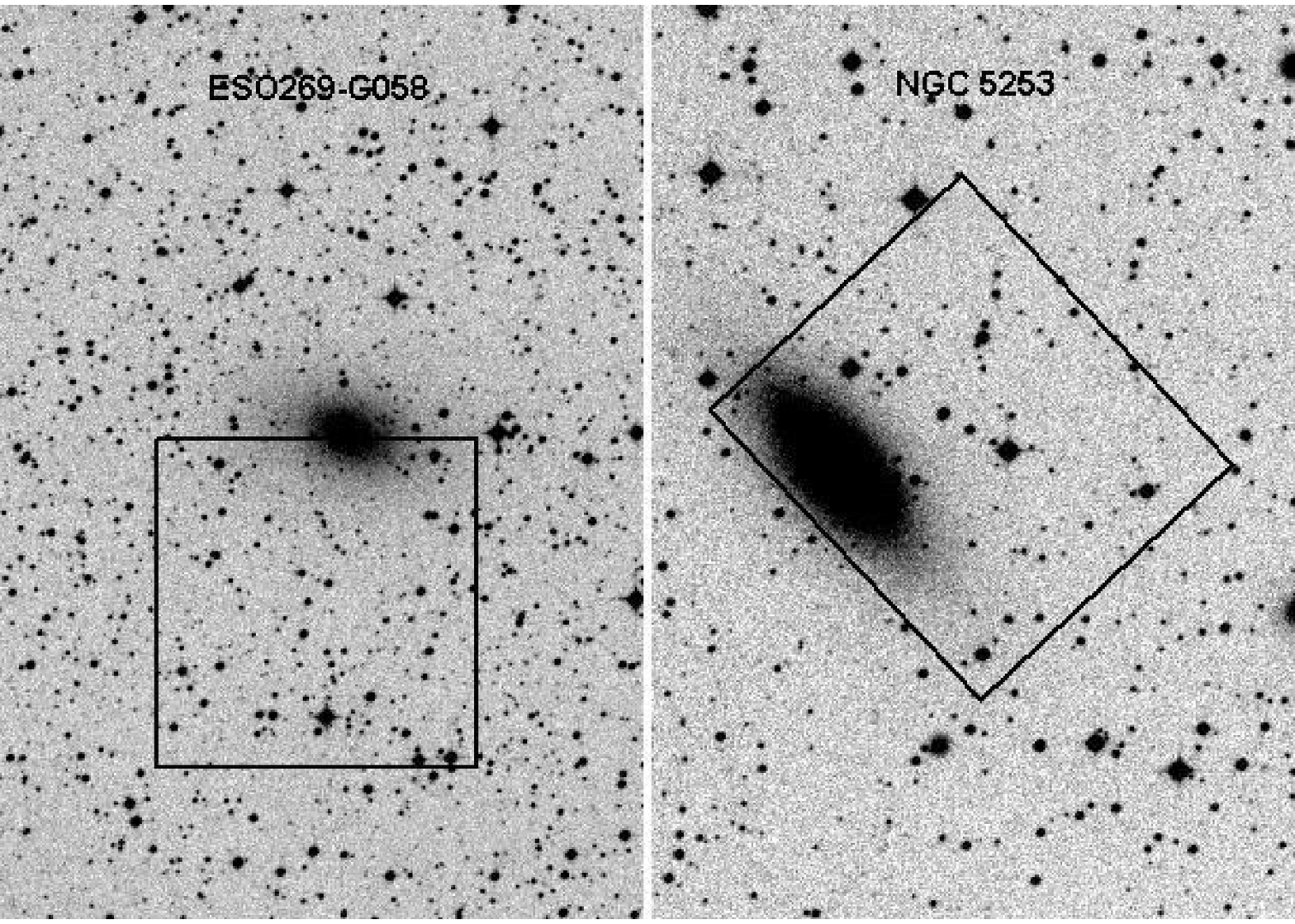}
\caption{Portions of the Digitized Sky Survey covering the $10 \times 15$ arcmin$^2$ 
area centered on each galaxy. The fields observed with GMOS are indicated. North 
is at the top, and east is to the left.}
\end{figure}

\clearpage
\begin{figure}
\figurenum{2}
\epsscale{0.75}
\plotone{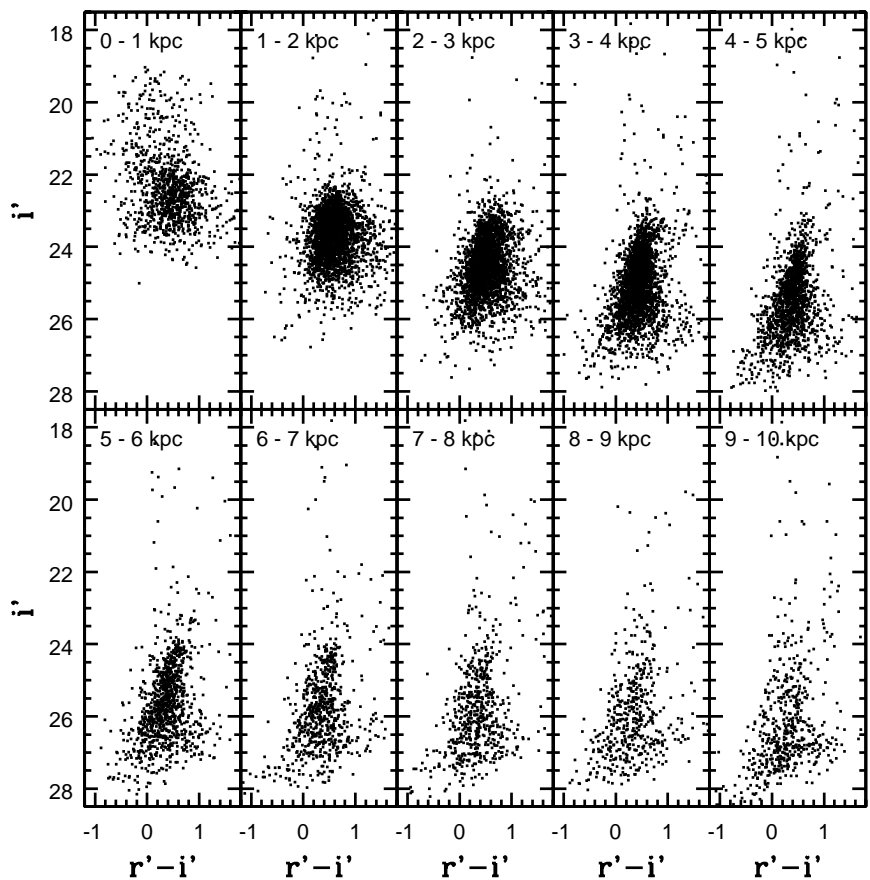}
\caption{The $(i', r'-i')$ CMDs of NGC 5253 and its surroundings. As in all subsequent 
figures, the distances listed in each panel are measured along the major 
axis of the galaxy. Note that RGB stars dominate the CMDs when R$_{GC}$ is between 
3 and 8 kpc. The CMDs are dominated by foreground stars and 
background galaxies when R$_{GC} > 8$ kpc.}
\end{figure}

\clearpage
\begin{figure}
\figurenum{3}
\epsscale{0.75}
\plotone{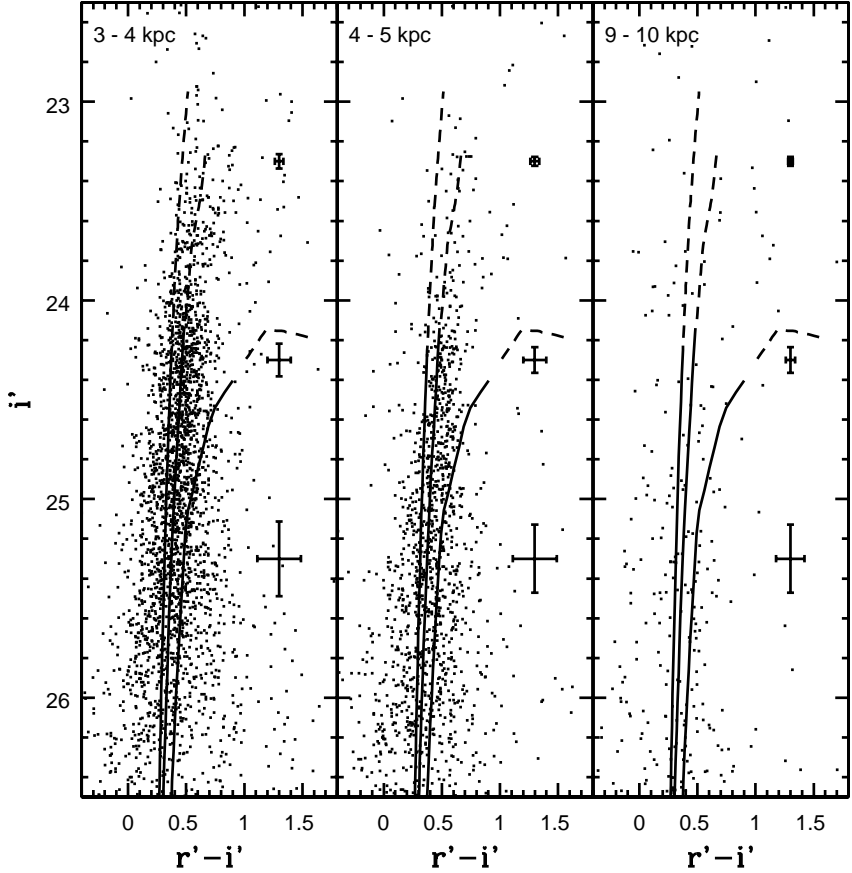}
\caption{The $(i', r'-i')$ CMDs of sources in three distance intervals near NGC 5253. The 
error bars show the uncertainties in the photometry predicted from the artificial star 
experiments. Isochrones from Girardi et al. (2004) for an age of 10 Gyr and Z = 0.0001, 
0.001, and 0.004 are also shown. The solid lines show evolution on the first ascent giant 
branch, while the dashed lines show the portion of the AGB where stellar luminosities 
exceed that of the RGB-tip. Note that (1) the colors of the RGB stars 
in the left hand and middle panels are consistent with Z $\sim 0.001$, (2) random 
errors in the photometry can account for much of the width of the RGB, and (3) a 
number of objects fall above the locus of AGB-tip values in the 3 -- 4 kpc interval, 
suggesting that a population of stars with ages $< 10$ Gyr might be present.}
\end{figure}

\clearpage
\begin{figure}
\figurenum{4}
\epsscale{0.75}
\plotone{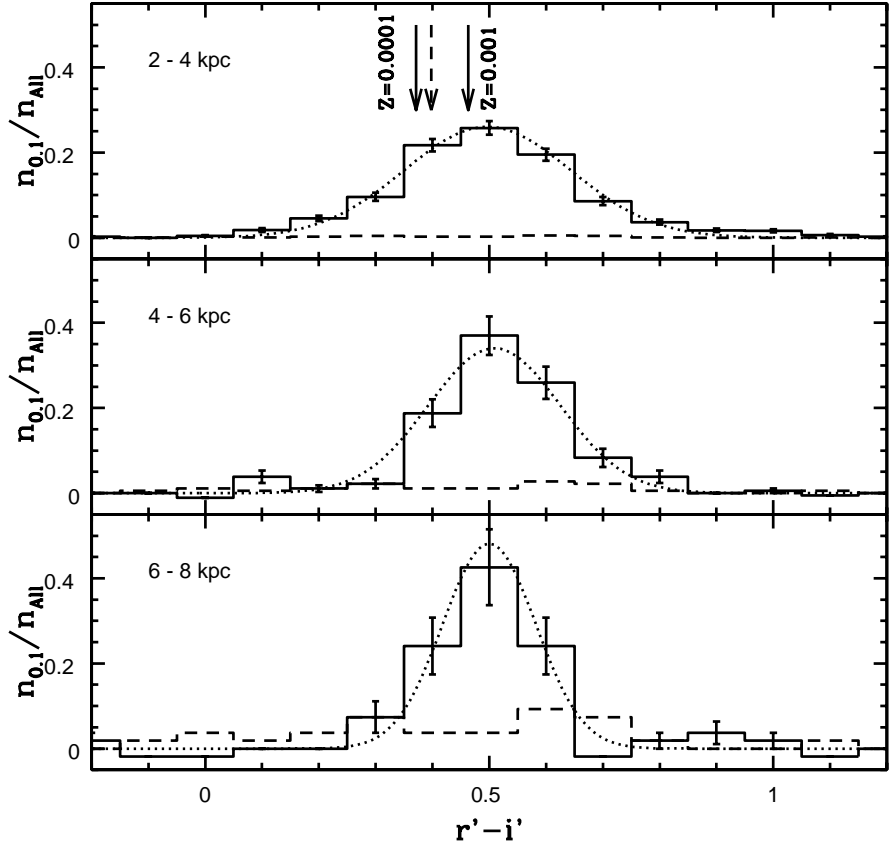}
\caption{The histogram distribution of $r'-i'$ colors for objects in NGC 5253 with $i'$ 
between 24.1 and 24.5 are shown as solid lines. n$_{0.1}$ is the number of stars per 0.1 
magnitude interval in $r'-i'$, while n$_{All}$ is the total number of stars in the 
brightness interval. The distributions have been corrected for contamination from 
foreground stars and background galaxies using the procedure described in the text. The 
solid arrows mark the color of the RGB at $i' = 24.3$ as 
predicted from the 10 Gyr Girardi et al. (2004) isochrones; 
a distance modulus $\mu_0 = 27.5$ and foreground extinction A$_B = 0.24$ have been 
assumed. The effect of age on the metallicity determination is demonstrated with the 
dashed arrow, which marks the color predicted for a Z = 0.001 with an age of 
2 Gyr. The dashed lines show the histogram distribution of objects in the 8 -- 
10 kpc interval, which is the region used to monitor contamination from foreground 
stars and background galaxies. These distributions have been scaled to show the 
fractional contribution made by contaminating objects to the total number of objects in 
each NGC 5253 distance interval. The dotted lines are Gaussians with dispersions 
computed from the differences between the known and measured brightnesses of 
artificial stars; these show the color distribution expected from an idealized 
RGB component that has no intrinsic color dispersion but is broadened by 
photometric errors. The excellent agreement between the observed histograms and the 
Gaussian distributions indicate that stars with metallicities that differ from the mean 
by more than a few tenths of a dex occur in only very small numbers in NGC 5253.} 
\end{figure}

\clearpage
\begin{figure}
\figurenum{5}
\epsscale{0.75}
\plotone{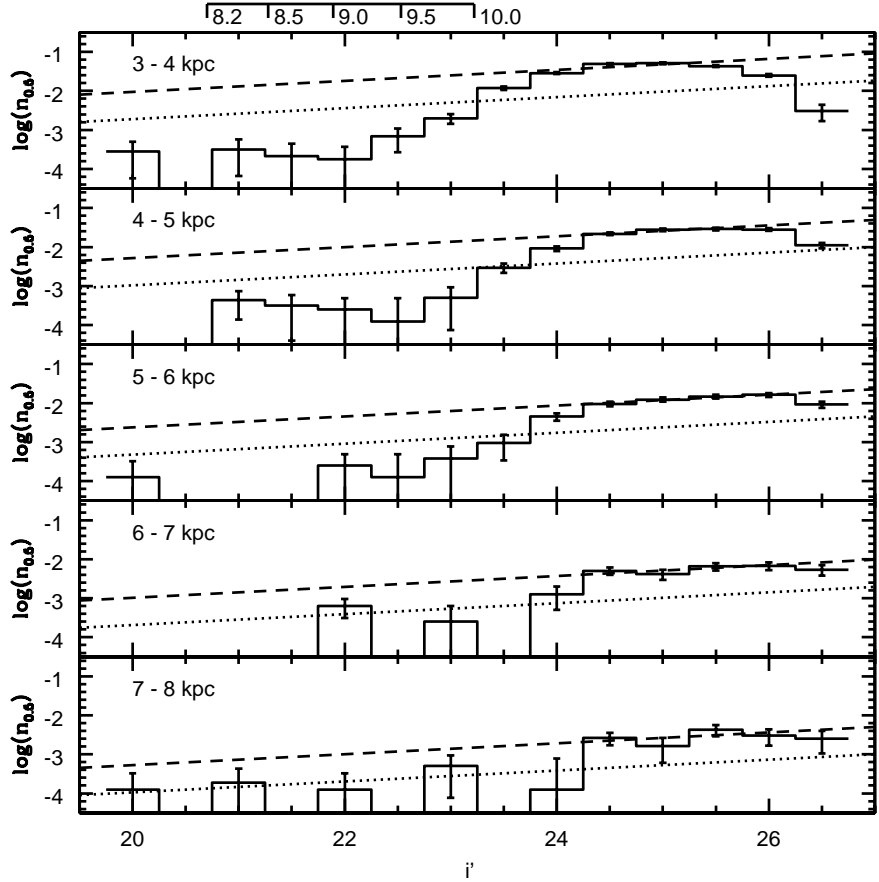}
\caption{The $i'$ LFs of stars with $r'-i'$ between -0.2 and 1.2, corrected for 
contamination from foreground stars and background galaxies using the procedure 
described in the text. n$_{0.5}$ is the number of stars per 0.5 magnitude 
interval arcsec$^{-2}$. The AGB-tip brightnesses predicted for populations 
with ages log(t$_{yr}$) = 8.2, 8.5, 9.0, 9.5, and 10.0 from the Girardi et al. (2004) 
Z = 0.001 isochrones are indicated at the top of the figure. 
The dashed lines are power-laws with an exponent that 
was computed from the $i' = 24.5$, 25.0, and 25.5 bins in the mean LF of 
objects with R$_{GC}$ between 4 and 8 kpc. The y-intercept of the result has been shifted 
so that it matches the counts in the $i' = 24.5$, 25.0, and 25.5 bins in each interval. 
The dotted lines show the expected trend for an old AGB population 
originating from the dominant RGB component, using the procedure 
described in the text. The number counts depart from the dashed line when 
$i' \leq 24.0$ due to the drop in stellar density associated with the RGB-tip.}
\end{figure}

\clearpage
\begin{figure}
\figurenum{6}
\epsscale{0.75}
\plotone{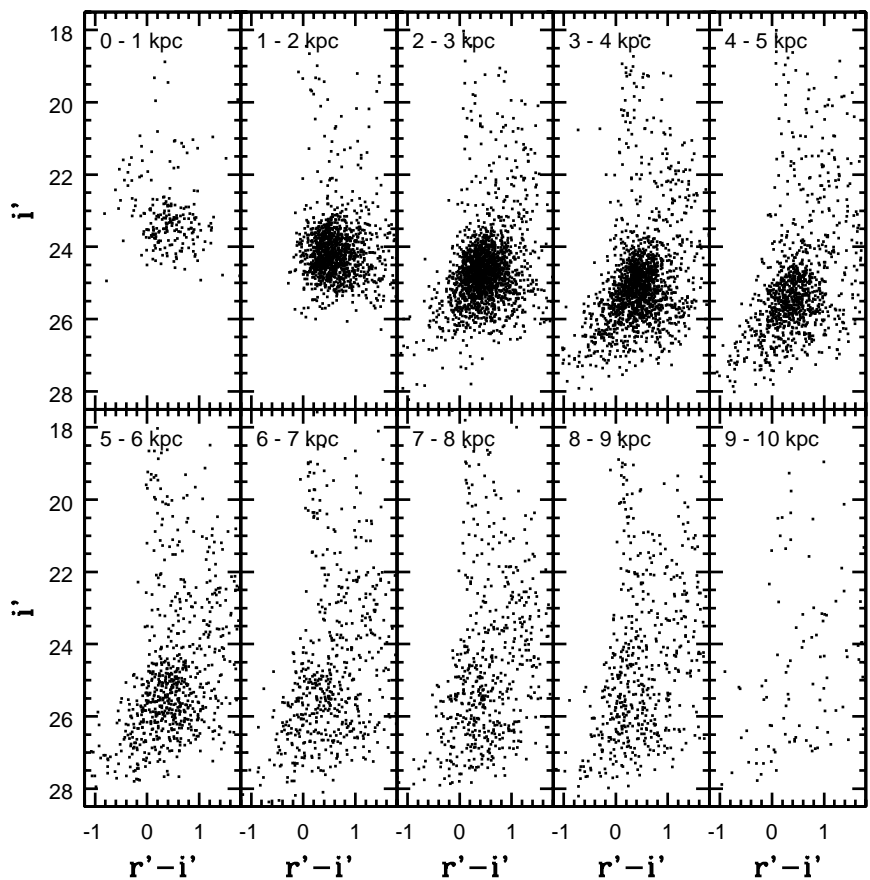}
\caption{The $(i', r'-i')$ CMDs of ESO269--G058 and its surroundings. 
The objects with $r'-i' < 0$ and $i' \sim 22$ in the 0 -- 1 kpc CMD are probably 
compact young star clusters. RGB stars dominate the CMDs of objects with R$_{GC}$ between 
2 and 6 kpc. When R$_{GC} > 6$ kpc the CMDs are dominated by foreground stars.}
\end{figure}

\clearpage
\begin{figure}
\figurenum{7}
\epsscale{0.75}
\plotone{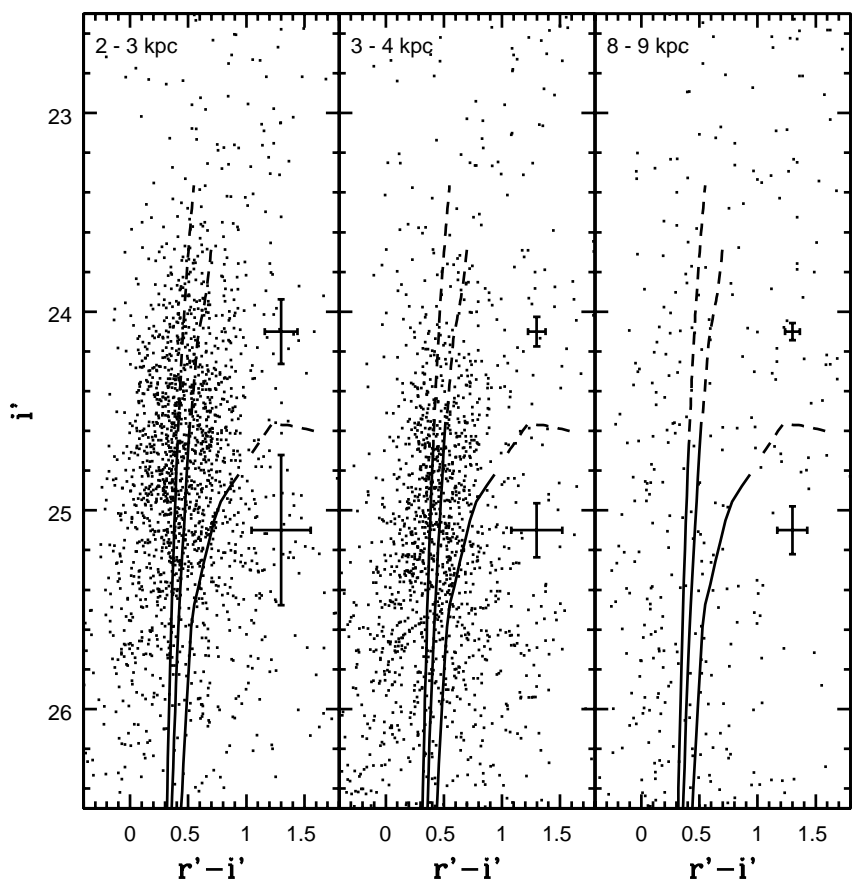}
\caption{The same as Figure 3, but showing the CMDs of three regions in and around 
ESO269--G058. The 8 -- 9 kpc CMD is dominated by foreground stars and background 
galaxies. Note that (1) the RGB of ESO269--G058 is broader than that of NGC 
5253 in Figure 3, (2) the locus of RGB stars is consistent with Z $\sim 0.001$, (3) 
a number of objects in the 2 -- 3 kpc interval fall above the 
envelope of AGB-tip values, suggesting that stars with ages 
$< 10$ Gyr may be present, and (4) the width of the RGB is roughly 
consistent with that expected from random photometric errors.}
\end{figure}

\clearpage
\begin{figure}
\figurenum{8}
\epsscale{0.75}
\plotone{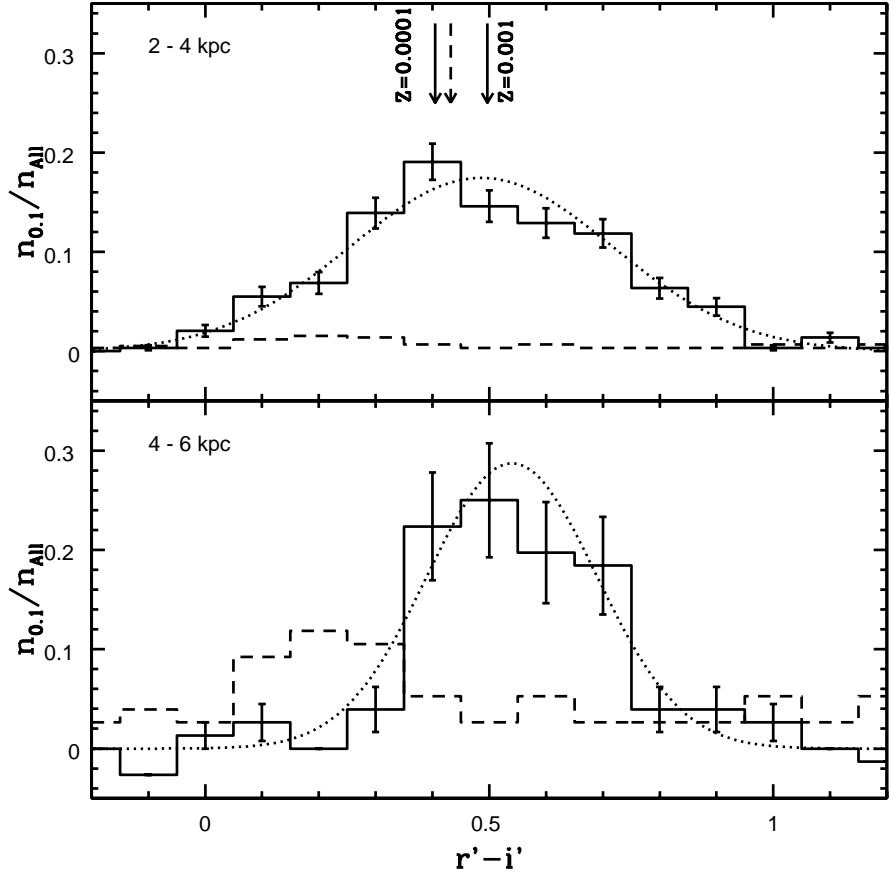}
\caption{The histogram distributions of $r'-i'$ colors for objects with $i'$ between 
24.7 and 25.1 in ESO269--G058. n$_{0.1}$ is the number of stars per 0.1 magnitude 
$r'-i'$ interval, while n$_{All}$ is the total number of stars in this color 
interval. The distributions have been corrected for contamination from foreground 
stars and background galaxies using the procedure described in the text. The arrows 
mark the color of the RGB at $i' = 24.9$ as predicted by the 10 Gyr Girardi et al. (2004) 
isochrones; a distance modulus $\mu_0 = 27.9$ and foreground extinction A$_B = 0.46$ 
have been assumed. The impact of age on the metallicity determination is demonstrated 
with the dashed arrow, which shows the color predicted for a Z = 0.001 system with 
an age of 2 Gyr. The dashed lines show the color distribution of objects in the 7 -- 
9 kpc interval, which is the distance interval that is used to track contamination from 
foreground stars and background galaxies. This distribution has been scaled to show the 
fractional contribution made by contaminating objects to the number of 
sources in each ESO269--G058 distance interval. The dotted lines are Gaussians with 
dispersions computed from the differences between the actual and measured brightnesses 
of artificial stars. These are the $r'-i'$ distribution expected from an idealized 
RGB with no intrinsic color dispersion but that is broadened by photometric errors. 
It is evident that photometric errors are the main source of broadening along 
the color axis amongst RGB stars in ESO269--G058.}
\end{figure}

\clearpage
\begin{figure}
\figurenum{9}
\epsscale{0.75}
\plotone{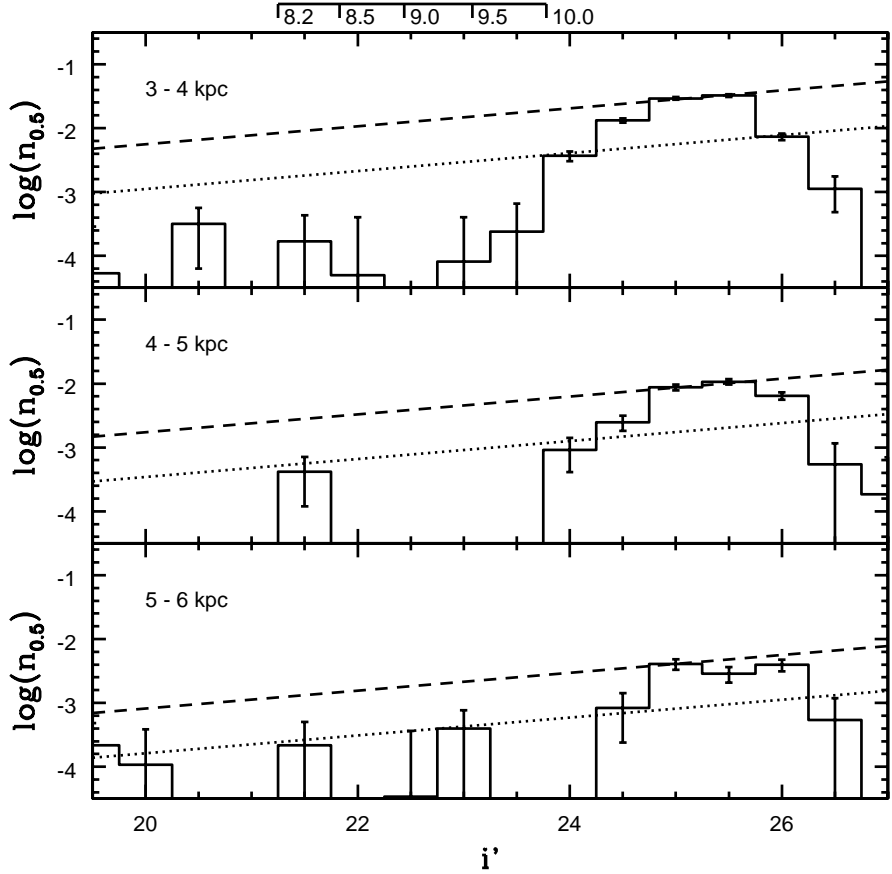}
\caption{The $i'$ LFs of stars with $r'-i'$ between -0.2 and 1.2 in ESO269--G058. 
The LFs have been corrected for contamination from 
foreground stars and background galaxies using the procedure 
described in the text. n$_{0.5}$ is the number of stars per 0.5 magnitude 
interval arcsec$^{-2}$. The dashed lines are power-laws with the exponent 
computed from the NGC 5253 RGB, but with the y-intercept shifted to match the 
counts in the $i' = 25$ bin in each interval. The number counts depart from the dashed 
line when $i' \leq 24.5$ due to the drop in stellar density 
caused by the RGB discontinuity. As in Figure 5, the dotted lines 
show the expected trend for an old AGB population originating from the 
dominant RGB component, while the AGB-tip brightnesses predicted from Z-=0.001 isochrones 
for various values of log(t$_{yr}$) are indicated above the top panel.} 
\end{figure}

\clearpage
\begin{figure}
\figurenum{10}
\epsscale{0.75}
\plotone{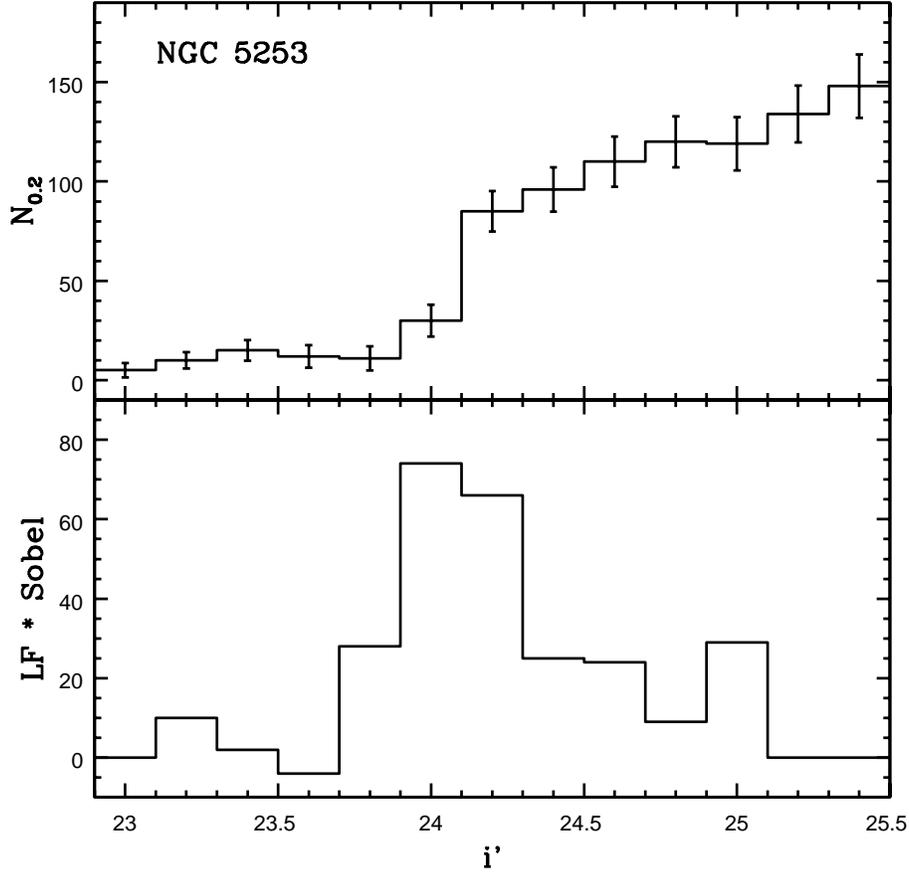}
\caption{Measuring the brightness of the RGB-tip in NGC 5253. The top panel 
shows the LF of stars with R$_{GC}$ between 4 and 6 kpc; N$_{0.2}$ is the number 
of stars per 0.2 magnitude bin in $i'$ with $r'-i'$ between --0.2 and 1.2, corrected for 
foreground stars and background galaxies based on the number counts of objects 
with R$_{GC}$ between 8 and 10 kpc. There is a distinct change in the LF when $i' 
= 24.1$, which is due to the onset of the RGB. The lower panel shows the result 
of convolving the LF in the top panel with a three point Sobel kernel. The 
convolved signal is highest in the $i' = 24.0$ and 24.2 bins, and the RGB-tip brightness 
deduced from the convolved signal is $i' = 24.09 \pm 0.01$.}
\end{figure}

\clearpage
\begin{figure}
\figurenum{11}
\epsscale{0.75}
\plotone{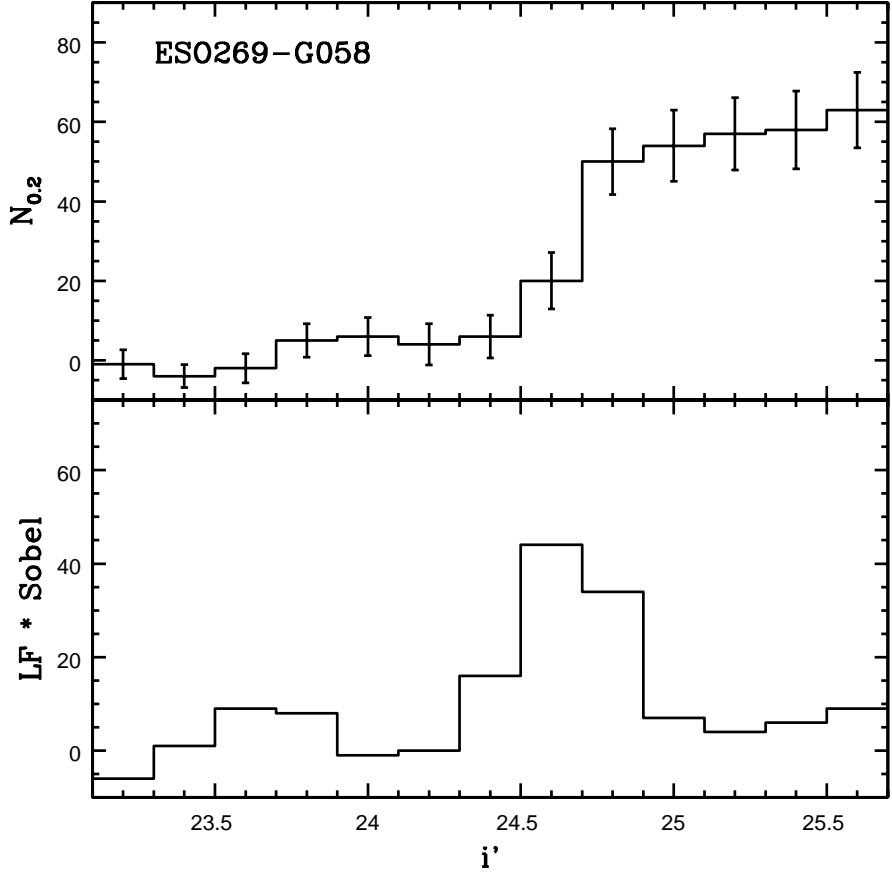}
\caption{Measuring the brightness of the RGB-tip in ESO269--G058. The top panel 
shows the LF of stars with R$_{GC}$ between 4 and 5 kpc; N$_{0.2}$ is the number 
of stars per 0.2 magnitude bin in $i'$ with $r'-i'$ between --0.2 and 1.2, corrected for 
foreground stars and background galaxies based on the number counts of objects 
with R$_{GC}$ between 7 and 9 kpc. The lower panel shows the result 
of convolving the LF in the top panel with a Sobel edge-detection kernel. The 
convolved signal is highest in the $i' = 24.4$, 24.6, and 24.8 bins, and the RGB-tip 
brightness deduced from these data is $i' = 24.64 \pm 0.10$.}
\end{figure}

\clearpage
\begin{figure}
\figurenum{A1}
\epsscale{0.75}
\plotone{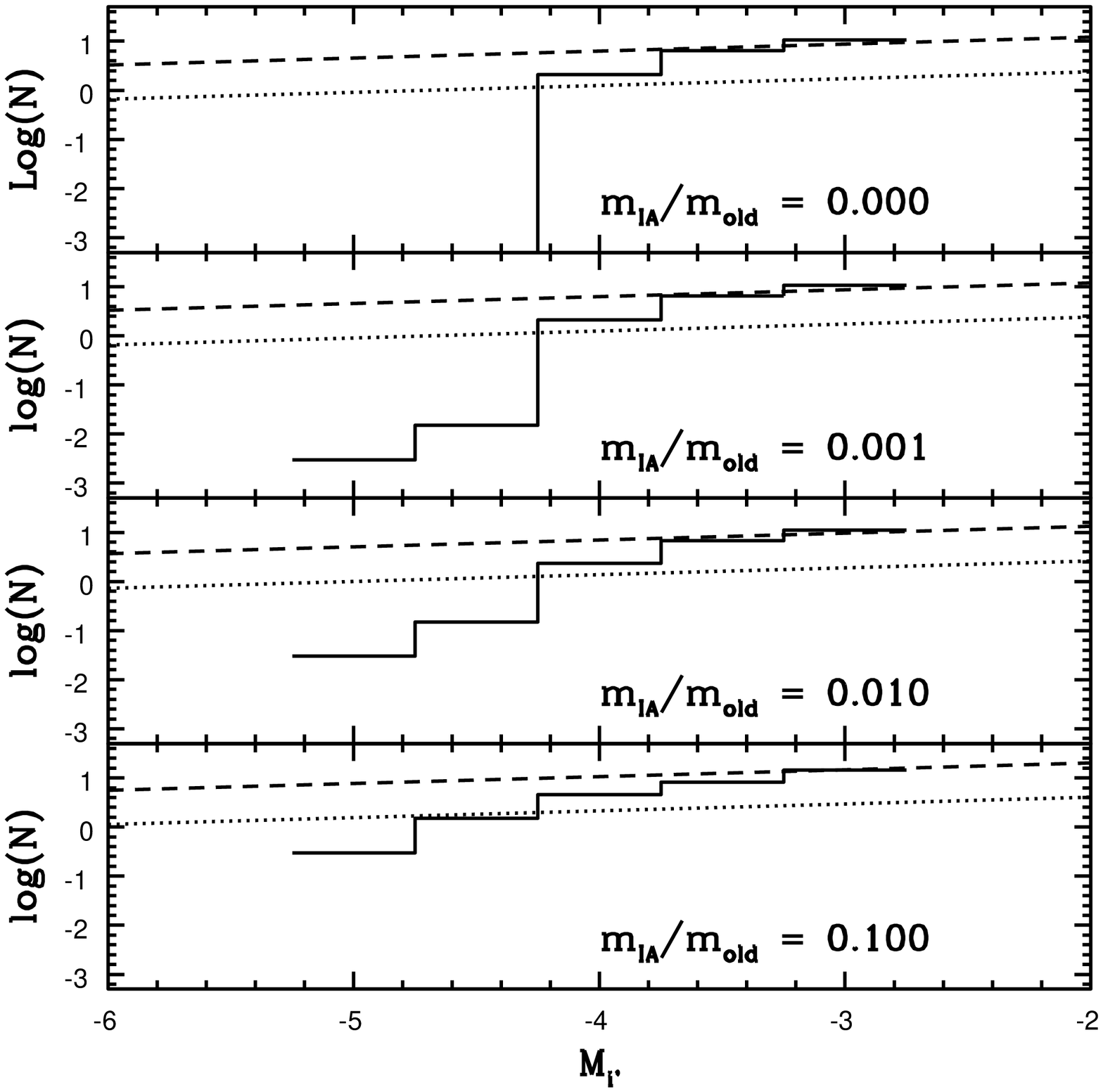}
\caption{Model LFs for various mixtures of an old and a 0.7 
Gyr population. The ratio of the mass of intermediate age to old stars is indicated in 
each panel. $N$ is the number of stars per 0.5 $i'$ magnitude interval in a system 
with a total mass of roughly 10$^5$ M$_{\odot}$. The dashed line shows the power law 
computed from the NGC 5253 RGB LF, but with the y-intercept shifted to match 
the number counts in the M$_{i'} = -3.5$ and --3 intervals. The dotted line shows this 
same relation, but shifted downwards by 0.7 dex, which provides an empirical 
measurement of the number of AGB stars in old stellar systems. Note that 
(1) the densities of bright AGB stars in the $i'$ LFs 
of NGC 5253 and ESO269--G058 at intermediate radii can be 
reproduced if the intermediate age population contributes only a few percent of the 
total stellar mass, and (2) if intermediate age stars contribute 10\% of the 
stellar mass then the discontinuity due to the RGB-tip is greatly suppressed.}
\end{figure}

\end{document}